\documentclass[]{article}
\usepackage{url}

\providecommand{\tightlist}{%
  \setlength{\itemsep}{0pt}\setlength{\parskip}{0pt}}

\usepackage{lmodern}
\usepackage{setspace}
\usepackage{amssymb,amsmath}
\usepackage{ifxetex,ifluatex}
\usepackage{fixltx2e} 
\ifnum 0\ifxetex 1\fi\ifluatex 1\fi=0 
  \usepackage[T1]{fontenc}
  \usepackage[utf8]{inputenc}
\else 
  \ifxetex
    \usepackage{mathspec}
    \usepackage{xltxtra,xunicode}
  \else
    \usepackage{fontspec}
  \fi
  \defaultfontfeatures{Mapping=tex-text,Scale=MatchLowercase}
  
\fi
\IfFileExists{upquote.sty}{\usepackage{upquote}}{}
\IfFileExists{microtype.sty}{%
\usepackage{microtype}
\UseMicrotypeSet[protrusion]{basicmath} 
}{}
\usepackage{longtable,booktabs}
\usepackage{graphicx}
\makeatletter
\def\maxwidth{\ifdim\Gin@nat@width>\linewidth\linewidth\else\Gin@nat@width\fi}
\def\maxheight{\ifdim\Gin@nat@height>\textheight\textheight\else\Gin@nat@height\fi}
\makeatother
\setkeys{Gin}{width=\maxwidth,height=\maxheight,keepaspectratio}
\ifxetex
  \usepackage[setpagesize=false, 
              unicode=false, 
              xetex]{hyperref}
\else
  \usepackage[unicode=true]{hyperref}
\fi
\hypersetup{breaklinks=true,
            bookmarks=true,
            pdfauthor={},
            pdftitle={Simulating Markov random fields with a conclique-based Gibbs sampler},
            colorlinks=true,
            citecolor=blue,
            urlcolor=blue,
            linkcolor=magenta,
            pdfborder={0 0 0}}
\title{\bf Simulating Markov random fields with a conclique-based Gibbs sampler}
\author{Andee Kaplan \\ Colorado State University \\ \texttt{}  and \\ Mark S. Kaiser \\ Iowa State University \\ \texttt{}  and \\ Soumendra N. Lahiri \\ Washington University in St.~Louis \\ \texttt{}  and \\ Daniel J. Nordman \\ Iowa State University \\ \texttt{} }
\date{}

\usepackage[toc,page]{appendix}
\usepackage{float}

\usepackage{amsthm}
\newtheorem{theorem}{Theorem}

\theoremstyle{definition}

\newcommand{\mbbeta}{\mbox{\boldmath $\beta$}}

\newcommand{\secspace}{~}

\newcommand{\hideFromPandoc}[1]{#1}
\hideFromPandoc{

}
\bigskip

\begin{document}

\def\spacingset#1{\renewcommand{\baselinestretch}%
{#1}\small\normalsize} \spacingset{1}

\maketitle

\begin{abstract}
For spatial and network data, we consider models formed from a Markov random field (MRF) structure and the specification of a conditional distribution for each observation. Fast simulation from such MRF models is often an important consideration, particularly when repeated generation of large numbers of data sets is required. However, a standard Gibbs strategy for simulating from MRF models involves single-site updates, performed with the conditional univariate distribution of each observation in a sequential manner, whereby a complete Gibbs iteration may become computationally involved even for moderate samples. As an alternative, we describe a general way to simulate from MRF models using Gibbs sampling with ``concliques'' (i.e., groups of non-neighboring observations). Compared to standard Gibbs sampling, this simulation scheme can be much faster by reducing Gibbs steps and independently updating all observations per conclique at once. The speed improvement depends on the number of concliques relative to the sample size for simulation, and order-of-magnitude speed increases are possible with many MRF models (e.g., having appropriately bounded neighborhoods). We detail the simulation method, establish its validity, and assess its computational performance through numerical studies, where speed advantages are shown for several spatial and network examples.
\end{abstract}
\noindent%

\newpage
\spacingset{1.45} 

\hypertarget{introduction}{%
\section{Introduction}\label{introduction}}

For modeling dependent data, conditionally specified models can be formulated on the basis of an underlying Markov random field (MRF) structure. This approach involves specifying a full conditional distribution for each observation, which often depends functionally on other (neighboring) observations in the conditional model statement (cf.~Besag \protect\hyperlink{ref-besag1974spatial}{1974}). Model formulation in this conditional-componentwise fashion provides an alternative to directly specifying a full joint data distribution. Such MRF models have become popular for spatially-dependent data (Cressie \protect\hyperlink{ref-cressie1993statistics}{1993}; Kaiser and Cressie \protect\hyperlink{ref-kaiser2000construction}{2000}), image segmentation (Zhang, Brady, and Smith \protect\hyperlink{ref-zhang2001segmentation}{2001}) and computer vision (Li \protect\hyperlink{ref-li2012markov}{2012}) among other applications including the analysis of networks (cf.~Strauss and Ikeda \protect\hyperlink{ref-strauss1990pseudolikelihood}{1990}; Hoff, Raftery, and Handcock \protect\hyperlink{ref-hoff2002latent}{2002}; Casleton, Nordman, and Kaiser \protect\hyperlink{ref-casleton2017local}{2017}). In addition to supporting model formulation, another pleasing aspect of MRF model specification is that univariate full conditional distributions fit naturally within the Gibbs sampling framework for simulating data for use in model assessment and Monte Carlo testing.

Accordingly, a dominant strategy for sampling from a MRF model involves a sequential update strategy with a Gibbs sampler whereby each observation in the field is simulated individually from its conditional distribution given all other current observational values (Besag, York, and Mollie \protect\hyperlink{ref-besag1991bayesian}{1991}). While simple to design in principle, single-site Gibbs updating can be slow, as each complete Gibbs iteration requires the same number of updates as there are data points. Consequently, even for relatively small data sets (e.g., a few hundreds of spatial points), there can be substantial time investments in just one run of the standard Gibbs sampler. These time investments are further compounded by the need for multiple iterations of this sampler in order to create a large collection of simulated data sets, as potentially required for ensuring appropriate mixing of the sampler (e.g., burn-in) and for adequately establishing some Monte Carlo approximation of interest (e.g., approximating a sampling distribution).

In this paper, we describe a simple and fast scheme for sampling from general MRF models in a manner that exploits conditional independence in such models among subcollections of non-neighboring observations called \emph{concliques}. ``Concliques'' provide a type of converse to ``cliques,'' where the latter are commonly encountered with MRFs as singletons or as sets of locations that are all mutual neighbors (Hammersley and Clifford \protect\hyperlink{ref-hammersley1971markov}{1971}). Kaiser, Lahiri, and Nordman (\protect\hyperlink{ref-kaiser2012goodness}{2012}) (hereafter {[}KLN{]}) introduced concliques to develop goodness-of-fit tests for spatial MRF models. However, apart from model assessment, the notion of concliques is shown here to have implications for potentially fast Gibbs sampling of MRFs. Namely, all MRF models admit concliques where number of concliques is never larger than, and in fact often much smaller than, the number of observations in the model. Consequently, we use concliques to establish a formal simulation method that applies under mild conditions to any conditionally specified MRF model and uses simultaneous updates (i.e., per conclique) to improve computational speeds. When the conclique number is smaller than the sample size for simulation, then for generating large collections of spatial data sets, the conclique-based approach can be computationally much more efficient than single-site Gibbs updating, while maintaining similar rates of chain mixing. Similarly to standard Gibbs sampling, the conclique-based strategy is also more generally applicable than alternative approaches for simulating from MRF models, such as those mentioned in Section \ref{other-simulation-approaches}.

In Section \ref{mrf-models-and-concliques}, we present some background about MRF models and concliques. Section \ref{conclique-based-gibbs-sampling} then describes the conclique-based Gibbs sampling approach, along with providing its theoretical justification and other ergodicity properties. Section \ref{numerical-comparisons-motivated-by-spatial-bootstrap} summarizes a numerical study of speed and convergence compared to standard sequential Gibbs sampling. In Section \ref{illustrations-with-networks}, we provide illustrations of the method for simulating networks. Concluding remarks are offered in Section\secspace\ref{concluding-remarks}, and the supplement contains further supporting theoretical and algorithmic results.

\hypertarget{other-simulation-approaches}{%
\subsection{Other simulation approaches}\label{other-simulation-approaches}}

We end this section with a brief overview of other simulation approaches for MRF models. While a joint data distribution, at least in theory, may be constructed from conditional distributions in a MRF specification, the normalizing terms involved are often intractable (cf.~Kaiser and Cressie \protect\hyperlink{ref-kaiser2000construction}{2000}). This motivates traditional use of a sequential Gibbs sampler based on individual conditional distributions, to which our proposed conclique-based Gibbs sampler is meant to be a computationally faster alternative.

Block Gibbs sampling with auxiliary variables, such as the Swendsen-Wang algorithm (Swendsen and Wang \protect\hyperlink{ref-swendsen1987nonuniversal}{1987}; Besag and Green \protect\hyperlink{ref-besag1993spatial}{1993}) and partial decoupling strategies (Higdon \protect\hyperlink{ref-higdon1994spatial}{1994}, \protect\hyperlink{ref-higdon1998auxiliary}{1998}), have been shown to improve chain mixing when attemping to sample from binary/multinomial MRF specifications close to criticality (i.e., where extreme dependence parameters induce near model-degeneracy (cf.~Kaiser and Caragea \protect\hyperlink{ref-kaiser2009exploring}{2009})). In such cases, single-site Gibbs samplers do not visit the potential data outcomes effectively, while instead chains based on block updates from auxiliary variables tend to mix more efficiently (cf.~Higdon \protect\hyperlink{ref-higdon1998auxiliary}{1998}). To be clear, the proposed conclique-based sampler is expected to share the same mixing weaknesses as the standard Gibbs sampler in such settings (i.e., only move faster through Gibbs iterations) and does not replace auxiliary variables. At the same time, in a general MRF application, standard Gibbs sampling can be superior to auxiliary variable approaches, where the latter can experience mixing slow-downs in presence of large-scale mean structures and face computational burdens in needing to determine, at each iteration, blocks or connected sets of observations to be updated (Hurn \protect\hyperlink{ref-hurn1997difficulties}{1997}; Higdon \protect\hyperlink{ref-higdon1998auxiliary}{1998}); see also Section \ref{a-numerical-study-of-simulation-efficacy} for a numerical comparison. The conclique-based sampler intends to provide speed advantages in similar general settings.

Simulation alternatives also exist to Gibbs sampling. Through chain coupling (Propp and Wilson \protect\hyperlink{ref-propp1996exact}{1996}), perfect sampling can apply for simulating from a MRF specification (cf.~Møller \protect\hyperlink{ref-moller1999perfect}{1999}), which has received particular consideration for generating lattice data from certain autologistic models (Friel and Pettitt \protect\hyperlink{ref-friel2004likelihood}{2004}; Hughes, Haran, and Caragea \protect\hyperlink{ref-hughes2011autologistic}{2011}; Hughes \protect\hyperlink{ref-hughes2014ngspatial}{2014}). But, due to the method's intricacies, perfect sampling does generally require more effort to set up than Gibbs sampling, as there is no exact rule for chain coupling. Additionally, perfect sampling also imposes some monotonicity requirements on conditional distributions which are not required in Gibbs sampling (Møller \protect\hyperlink{ref-moller1999perfect}{1999}). When considering Gaussian MRF models, several further possibilities exist for data simulation, including direct sampling and circulant embedding (Rue \protect\hyperlink{ref-rue2001fast}{2001}; Rue and Held \protect\hyperlink{ref-rue2005gaussian}{2005}; Møller and Waagepetersen \protect\hyperlink{ref-moller2003statistical}{2003}; Davies and Bryant \protect\hyperlink{ref-davies2013circulant}{2013}). However, even for Gaussian MRF models, the simplicity of the Gibbs sampler is attractive. Ultimately, for MRF specifications, Gibbs sampling plays a natural role in simulation from a broad variety of discrete and continuous data structures on both regular and irregular lattices (e.g., spatial or network data), where concliques may provide a beneficial tool in simulation.

\hypertarget{mrf-models-and-concliques}{%
\section{MRF models and concliques}\label{mrf-models-and-concliques}}

\hypertarget{mrf-formulation}{%
\subsection{MRF formulation}\label{mrf-formulation}}

We introduce some notation for MRF models using, for concreteness, a description typical in an applied spatial context. Let \(\{\boldsymbol s_i:i=1,\dots,n\}\) represent a set of locations, generically indexed in some Euclidean space (e.g., \(\mathbb{R}^2\)), and let \(\{Y(\boldsymbol s_i):i=1,\dots,n\}\) denote a corresponding collection of indexed univariate random variables. A MRF formulation commonly involves specifying a neighborhood for each location \(\boldsymbol s_i\), which consists of locations on which the full conditional distribution of \(Y(\boldsymbol s_i)\) is functionally dependent. Let \(f_i\) denote the conditional density (or mass) function of \(Y(\boldsymbol s_i)\) given
all other observations \(\{Y(\boldsymbol s_j)=y(\boldsymbol s_j): j \neq i\}\), noting that a common density form (\(f_i = f\)) may also be applied. Additionally, let
\(\mathcal{N}_i \equiv \{ \boldsymbol s_j: i \neq j \text{ and } f_i \text{ depends functionally on } y(\boldsymbol s_j)\}\) represent the neighborhood for location \(\boldsymbol s_i\) and state a corresponding set of neighborhood observations as
\(\boldsymbol y(\mathcal{N}_i)\equiv \{y(\boldsymbol s_j):\boldsymbol s_j \in \mathcal{N}_i\}\). Under a defining MRF assumption, it holds that
\begin{equation}
\label{eqn:1}
f_i(y(\boldsymbol s_i)| \{y(\boldsymbol s_j): j \neq i\}) = f_i(y(\boldsymbol s_i)| \boldsymbol y(\mathcal{N}_i)).
\end{equation}
The model is formed by prescribing a full conditional density (\ref{eqn:1}) for each observation \(i=1,\dots,n\). We shall assume that a valid joint distribution exists for
\(\{Y(\boldsymbol s_i), \dots, Y(\boldsymbol s_n)\}\) that corresponds to the conditionals specified in (\ref{eqn:1}). Arnold, Castillo, and Sarabia (\protect\hyperlink{ref-arnold2001conditionally}{2001}) provide conditions necessary for such a joint to exist, while Kaiser and Cressie (\protect\hyperlink{ref-kaiser2000construction}{2000}) describe conditions under which a joint may be constructed on the basis of the specified conditionals. In developing a MRF model for data, model diagnostics may be performed with goodness-of-fit tests (cf.~{[}KLN{]} and the data application in the supplement), while Kaiser and Nordman (\protect\hyperlink{ref-kaiser2012blockwise}{2012}) describe an approach for testing neighborhood structures with spatial lattice data.

One common example of conditional densities in a MRF specification (\ref{eqn:1}) involves an exponential family form given by
\begin{equation}
\label{eqn:2}
f_i(y(\boldsymbol s_i)|\boldsymbol y(\mathcal{N}_i), \boldsymbol \theta) = \exp\left[A_{i}(\boldsymbol y(\mathcal{N}_i))y(\boldsymbol s_i) - B_i(\boldsymbol y(\mathcal{N}_i)) + C(y(\boldsymbol s_i))\right],
\end{equation}
where \(A_{i}(\cdot)\) is a natural parameter function, \(B_i(\cdot)\) is a function of \(\boldsymbol y(\mathcal{N}_i)\) only through \(A_{i}(\cdot)\), and \(C(\cdot)\) is a known function. Under an assumption of pairwise-only dependence (or cliques of at most size two), Besag (\protect\hyperlink{ref-besag1974spatial}{1974}) showed a necessary form in (\ref{eqn:2}) as \(A_i(\boldsymbol y(\mathcal{N}_i)) = \alpha_i + \sum_{i=1}^n \eta_{i,j}y(\boldsymbol s_j)\) with parameters \(\alpha_i\), \(\eta_{i,i}=0\), \(\eta_{i,j}=\eta_{j,i}\) and \(\eta_{i,j} = 0\) unless \(\boldsymbol s_j \in \mathcal{N}_i\). For many models, a useful parametrization is given by \(A_i(\boldsymbol y(\mathcal{N}_i)) = \tau^{-1}(\kappa_i) + \sum_{\boldsymbol s_j \in \mathcal{N}_i} \eta_{i,j}\{y(\boldsymbol s_j) - \kappa_j\}\), with dependence parameters \(\eta_{i,j} = \eta_{j,i}\), a large scale parameter \(\kappa_i\), and a function \(\tau^{-1}(\cdot)\) that maps expected values to natural parameters; see Kaiser, Caragea, and Furukawa (\protect\hyperlink{ref-kaiser2012centered}{2012}).

Numerical studies in Sections \ref{conclique-based-gibbs-sampling}-\ref{illustrations-with-networks} consider some MRF examples (\ref{eqn:1})-(\ref{eqn:2}) in more detail. Neighborhoods \(\mathcal{N}_i\) in a MRF structure are flexible and, for describing concliques in Section \ref{concliques}, may be treated separately from the kind of distribution used in a conditional specification (\ref{eqn:1}).

\hypertarget{concliques}{%
\subsection{Concliques}\label{concliques}}

The MRF model (\ref{eqn:1}) again involves, for each observation \(Y(\boldsymbol s_i)\), a conditional distribution \(f_i\) that depends on observations \(\boldsymbol y(\mathcal{N}_i)\) in a neighborhood \(\mathcal{N}_i\) of location \(\boldsymbol s_i\). From this model formulation, a \emph{conclique} defined by {[}KLN{]} is a singleton set or a set of locations such that no location in the set is a neighbor of any other location in the set. Any MRF specification always admits a collection of concliques, say \(\mathcal{C}_1, \dots, \mathcal{C}_Q\), that partition the available spatial locations as
\(\cup_{i=1}^Q \mathcal{C}_i = \{\boldsymbol s_1,\ldots,\boldsymbol s_n\}\) with \(\mathcal{C}_i \cap \mathcal{C}_j=\emptyset\) for \(i\neq j\). Some examples of concliques are considered next, while Section \ref{finding-concliques} describes a device for quantifying the number \(Q\) of concliques needed for a specified MRF model and provides guidance for determining concliques.

\begin{figure}
\centering
\includegraphics{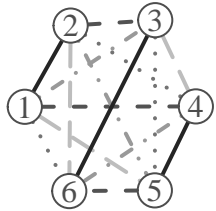}
\caption{\label{fig:graph-conc-plot}\label{fig:graph-conc}Five concliques (indicated by shade and line type) partition the \({6 \choose 2}\) edge variables in a graph with \(V=6\) vertices under ``incidence'' neighborhoods.}
\end{figure}

From spatial data modeling, three standard neighborhood structures with observations on a regular lattice are given by two-, four-, and eight-nearest neighbors (Besag \protect\hyperlink{ref-besag1974spatial}{1974}). As depicted in Figure \ref{fig:concliques}, a two-nearest neighborhood may be formed by two ``unilateral'' locations \(\mathcal{N}_i = \{\boldsymbol s_i + \boldsymbol h : \boldsymbol h = \pm (1, 0)\}\); a four-nearest neighborhood is comprised of locations in cardinal directions as \(\mathcal{N}_i = \{\boldsymbol s_i + \boldsymbol h : \boldsymbol h = \pm (0, 1), \pm (1, 0)\}\); and the eight-nearest neighbor neighborhood \(\mathcal{N}_i = \{\boldsymbol s_i + \boldsymbol h : \boldsymbol h = \pm (0, 1),\pm (1, 0), \pm (1, -1), \pm (1, 1)\}\) further includes neighboring diagonals. Consequently, it is possible to partition locations into two concliques under the two- or four-nearest neighborhood structures but into four concliques under the eight-nearest neighborhood, as indicated in Figure \ref{fig:concliques}. The prototypical types of neighborhoods and concliques given in Figure \ref{fig:concliques} are often considered in spatial illustrations to follow. For regular lattices, these particular concliques also correspond to the so-called coding sets of Besag (\protect\hyperlink{ref-besag1974spatial}{1974}), which were suggested in developing pseudo-likelihood estimation. The defining characteristic of concliques, however, allows identification of such sets in broader settings including graphs and networks (cf.~Sec.\secspace\ref{illustrations-with-networks}) as well as other irregular lattices. For example, as a small illustration of MRF neighborhoods/concliques for network data, consider the \(n \equiv V(V-1)/2\) possible edges in a simple graph with \(V\) vertices, where we associate a random variable \(Y(\boldsymbol{s}_i)\), \(i=1,\ldots,n\) with each edge ``marker'' \(\boldsymbol{s}_i = \{v_{i1}, v_{i2}\}\) defined by two graph vertices (say, \(v_{i1}, v_{i2}\)) for prescribing an edge's location. Commonly, a binary \(Y(\boldsymbol{s}_i)\) indicates the presence/absence of a ``random edge'' at location \(\boldsymbol{s}_i\) and a so-called ``incidence'' neighborhood \(\mathcal{N}_i = \{\boldsymbol {s}_j : \boldsymbol {s}_i \cap \boldsymbol {s}_j \neq \emptyset\}\) is given by other edge locations \(\boldsymbol {s}_j\) sharing a common node with \(\boldsymbol {s}_i\) (cf.~Frank and Strauss \protect\hyperlink{ref-frank1986markov}{1986}). In this case, concliques simply consist of collections of graph edges that share no nodes, as illustrated in Figure \ref{fig:graph-conc-plot} for a graph with \(V=6\) vertices. Section \ref{illustrations-with-networks} returns to examples of random networks in more detail.

\par
\begin{figure}[t]
\vspace*{.15cm}
\noindent
\begin{minipage}{.33\linewidth}
\centering
\underline{Two-nearest}   \\[-.5cm]
$$
\begin{array}{ccc}
 &\mathcal{N}_i&\\
\cdot&\cdot&\cdot\\
*&\boldsymbol s_i&*\\
\cdot&\cdot&\cdot\\
&&\\
\end{array} \quad \begin{array}{cccc}
\multicolumn{4}{c}{\mbox{{\it Concliques}}}\\
1&2&1&2\\
2&1&2&1\\
1&2&1&2\\
2&1&2&1\\
\end{array}
$$
\end{minipage}
\begin{minipage}{.33\linewidth}
\centering
\underline{Four-nearest}   \\[-.5cm]
$$
\begin{array}{ccc}
 &\mathcal{N}_i&\\
\cdot&*&\cdot\\
*&\boldsymbol s_i&*\\
\cdot&*&\cdot\\
&&\\
\end{array} \quad  \begin{array}{cccc}
\multicolumn{4}{c}{\mbox{{\it Concliques}}}\\
1&2&1&2\\
2&1&2&1\\
1&2&1&2\\
2&1&2&1\\
\end{array}
$$
\end{minipage}\begin{minipage}{.33\linewidth}
\centering
\underline{Eight-nearest} \\[-.5cm]
$$
\begin{array}{ccc}
&\mathcal{N}_i& \\
*&*&*\\
*&\boldsymbol s_i&*\\
*&*&*\\
&&\\
\end{array}\quad \begin{array}{cccc}
\multicolumn{4}{c}{\mbox{{\it Concliques}}}\\
1&2&1&2\\
3&4&3&4\\
1&2&1&2\\
3&4&3&4\\
\end{array}
$$
\end{minipage}
\caption{Illustration of  two-, four-, and eight-nearest neighborhoods $\mathcal{N}_i$ (neighbors of $\boldsymbol s_i$ denoted by $*$) and sets of concliques (represented by similar numbers that denote non-neighbors).
}
\label{fig:concliques}
\end{figure}
Note that any subdivision of a conclique necessarily results in subsets which are also concliques. In practice, we wish to identify a small number \(Q\) of concliques, which is valuable as the proposed simulation procedure requires one Gibbs step for each conclique. A minimal collection of concliques, or a so-called minimal conclique cover {[}cf.~KLN{]}, is achieved when the number \(Q\) of concliques is as small as possible. For example, minimal conclique covers have sizes \(Q=2\) and \(Q=4\), respectively, for the four- and eight-nearest neighbor spatial schemes (Figure \ref{fig:concliques}); in the network example above involving \(n \equiv V(V-1)/2\) edges in a graph with \(V\) vertices, minimal conclique covers exist of size \(Q = 2\lceil V/2 \rceil -1\) (cf.~Figure \ref{fig:graph-conc-plot}), where the supplement provides a construction. General guidance on determining concliques is given next, including some worst-case bounds for judging how many concliques may be required for a MRF model before performing an actual conclique determination.

\hypertarget{finding-concliques}{%
\subsection{Finding concliques}\label{finding-concliques}}

The MRF simulation approach to follow requires a one-time determination of concliques from conditional models (\ref{eqn:1}). Minimal conclique covers are ideal, but not crucial, for computational advantages over single-site Gibbs samplers. Speed improvements require that the number \(Q\) of concliques used be effectively smaller than the sample size \(n\) intended for simulation. For many MRF models, the latter feature often occurs with concliques found through simple procedures, including graph coloring algorithms, described below. However, without explicitly determining concliques, some basic bounds can be given to frame the potential size of \(Q\) relative to \(n\). As a first reference, note that \(Q \leq n\) always holds and the case \(Q=n\) corresponds to an extreme scenario for concliques (i.e., one observation \(\mathcal{C}_i=\{\boldsymbol{s}_i\}\) per conclique). Hence, in its worst case implementation, a conclique-based Gibbs sampler becomes a single-site sampler (i.e., same as the standard Gibbs strategy). Further, based on the sizes \(|\mathcal{N}_i|\) of neighborhoods \(\{\mathcal{N}_i\}_{i=1}^n\) from a given MRF model (\ref{eqn:1}), the number \(Q\) of concliques needed may also be usefully delimited, as follows, for comparison to \(n\): if neighborhood sizes are ordered as \(|\mathcal{N}_{(1)}| \leq |\mathcal{N}_{(2)}| \leq \cdots \leq |\mathcal{N}_{(n)}| \leq n-1\), then the number \(Q\) of concliques required under the model will be no more than \(\Delta_{n}\) for
\begin{equation}
\label{eqn:bound}
\Delta_n \equiv \max_{1 \leq i \leq n}\min\{|\mathcal{N}_{(n-i+1)}|+1, i  \} \;\leq\;
|\mathcal{N}_{(n)}|+1;
\end{equation}
this follows from a graph coloring bound described below. Hence, as one point of reference from (\ref{eqn:bound}), when the largest neighborhood size \(|\mathcal{N}_{(n)}|\) is smaller order than \(n\), as is common in many MRF model specifications (cf.~Sec \ref{concliques}), then the required number \(Q\) of concliques will be similarly smaller order than \(n\). However, from \(\Delta_n\) in (\ref{eqn:bound}), a relatively small number of concliques can also follow under a variety of possible neighborhood configurations, including cases where some neighborhoods are extremely large or entirely connected (\(|N_{(i)}|=n-1\)) as long as the number of such neighborhoods is appropriately limited (e.g., smaller order than \(n\)).

With regard to explicitly finding concliques, if simulation involves lattice data with a local and common neighborhood structure, then concliques may be determined from physical distance considerations (i.e., ``basic concliques'' described in {[}KLN{]}) so that the number \(Q\) of concliques is often no greater than the neighborhood size (cf.~concliques in Figure \ref{fig:concliques}). For finding concliques more generally, we may use algorithms for graph coloring (cf.~Jensen and Toft \protect\hyperlink{ref-jensen2011graph}{2011}): namely, finding the smallest (or chromatic) number of colors needed to color graph vertices in a way that no edge-connected vertices share the same color. In this way, the concept of concliques (or non-neighboring sites) translates to notion of non-adjacent (commonly colored) vertices in graph theory. While many sophisciated algorithms exist and could be used for graph coloring (cf.~Husfeldt \protect\hyperlink{ref-husfeldt2015graph}{2015} for a review), our experience is that greedy search algorithms often suffice for finding concliques/colorings. Such algorithms tend to be simple but effective at finding low numbers of colors (or small numbers \(Q\) of concliques), though without guarantees that a chromatic number of colors will be found. For example, a basic greedy search (known as the Welch-Powell algorithm) that moves through the sites \(\{\boldsymbol {s}_1,\ldots,\boldsymbol {s}_n\}\) labeled by decreasing neighborhood size \(|\mathcal{N}_{i}|\geq |\mathcal{N}_{i+1}|\), and adds each site to an existing conclique (otherwise starting a new conclique), is known to yield at most \(\Delta_n\) concliques as in (\ref{eqn:bound}) (cf.~Husfeldt \protect\hyperlink{ref-husfeldt2015graph}{2015}). While the latter algorithm can be simply and quickly used, we might also recommend DSatur (based on ``saturation degree'' from Brélaz (\protect\hyperlink{ref-brelaz1979new}{1979})), which is a well-known dynamic greedy coloring algorithm with portable R and python implementations (Hunziker \protect\hyperlink{ref-mapcoloring}{2017}; Novikov \protect\hyperlink{ref-novikovpyclustering}{2019}). Though only provably optimal when two colorings exist, DSatur is fast among coloring algorithms and known to broadly perform well at finding small numbers \(Q\) of graph colors (or concliques). For example, DSatur returns the minimal conqlique covers for the 4- and 8- nearest spatial neighborhoods in the examples of Figure \ref{fig:concliques} as well as finds the \(Q = 2 \lceil V/2\rceil - 1\) minimal conclique covers for the size \(n={V \choose 2}\) networks corresponding to Figure \ref{fig:graph-conc-plot} (i.e., \(V\) vertices and ``incidence'' neighborhoods). In contrast, the Welch-Powell algorithm typically fails to locate minimal covers (e.g., often returning 4 and 8, not 2 and 4, concliques under 4- and 8-nearest neighborhoods), though this aspect is again not ultimately important for speed improvements over standard Gibbs sampling in simulation when the bound \(\Delta_n\) in (\ref{eqn:bound}) has smaller order than \(n\).

\hypertarget{conclique-based-gibbs-sampling}{%
\section{Conclique-based Gibbs sampling}\label{conclique-based-gibbs-sampling}}

\hypertarget{method-of-simulation}{%
\subsection{Method of simulation}\label{method-of-simulation}}

To frame the simulation approach for MRFs to follow, we recall a result of {[}KLN{]} regarding concliques and conditional probability integral transforms. Let \(F_i\) denote the cumulative distribution function (cdf) for the conditional density \(f_i\) in (\ref{eqn:1}) of observation \(Y(\boldsymbol s_i)\), assuming this cdf is continuous for simplicity, and define a residual \(U(\boldsymbol s_i) = F_i(Y(\boldsymbol s_i)| \{Y(\boldsymbol s_j) : \boldsymbol s_j\in\mathcal{N}_i\})\) for location \(\boldsymbol s_i\) by substituting observations into the conditional cdf form. As shown by {[}KLN{]} in developing goodness-of-fit statistics, such residuals are iid Uniform\((0,1)\) distributed within each conclique: that is, for each \(j = 1, \dots, Q\), the collection \(\{U(\boldsymbol s_i) : \boldsymbol s_i \in \mathcal{C}_j\}\) of residuals provides a Uniform\((0,1)\) random sample. This result may be interpreted as a means to independently generate observations for an entire conclique \(\mathcal{C}_j\) given observations from other concliques \(\mathcal{C}_k\), \(k\neq j\): draw a random sample, say \(\{U^*(\boldsymbol s_i) : \boldsymbol s_i \in \mathcal{C}_j\}\), of Uniform\((0,1)\) variables and compute \(Y(\boldsymbol s_i) \equiv F_i^{-1}(U^*(\boldsymbol s_i))\), \(\boldsymbol s_i \in \mathcal{C}_j\), where any conditioning observation required in \(F_i\), for \(\boldsymbol s_i \in \mathcal{C}_j\), must belong to \(\mathcal{C}_k\) for some \(k\neq j\). For simulating from MRF models, a Gibbs sampler may then be formulated in an alternative fashion to the standard approach so that updates are conducted independently and simultaneously per conclique. An algorithm for this conclique-based Gibbs sampler (CGS) is presented next.

\textbf{CGS Algorithm:} Let \(Y^{(m)}(\boldsymbol s)\) denote the value of an observation at location \(\boldsymbol s\) at the \(m\)th sampling iteration, \(m=0,1,\ldots, M\), where \(M \geq 1\) is the desired number of iterations.
\begin{enumerate}
\def\labelenumi{\Alph{enumi}.}
\tightlist
\item
  Split intended locations \(\{\boldsymbol s_1,\ldots, \boldsymbol s_n\}\) into \(Q \geq 2\) disjoint concliques, \(\mathcal{C}_1, \ldots,\mathcal{C}_Q\).
\item
  Initialize values for observations \(\{Y^{(0)}(\boldsymbol s): \boldsymbol s \in \{\mathcal{C}_2, \dots, \mathcal{C}_Q\}\}\) outside conclique \(\mathcal{C}_1\).
\item
  For iteration \(m = 1, \dots, M\),
  \begin{enumerate}
  \def\labelenumii{\arabic{enumii}.}
  \tightlist
  \item
    Considering all locations \(\boldsymbol s_i \in \mathcal{C}_1\),
    sample \(\{Y^{(m)}(\boldsymbol s_i) : \boldsymbol s_i \in \mathcal{C}_1 \}\) by independently drawing \(Y^{(m)}(\boldsymbol s_i) \sim f_i(\cdot|\{Y^{(m-1)}(\boldsymbol s), \boldsymbol s \in \mathcal{N}_i\})\) from conditionals in (\ref{eqn:1}).
  \item
    Set \(\ell =2\).
  \item
    Considering all locations \(\boldsymbol s_i \in \mathcal{C}_\ell\), sample \(\{Y^{(m)}(\boldsymbol s_i): \boldsymbol s_i \in \mathcal{C}_\ell\}\) by independently drawing \(Y^{(m)}(\boldsymbol s_i) \sim f_i(\cdot|\boldsymbol y_\ell^{(m)}(\mathcal{N}_i))\) with conditioning observations
    \[
    \boldsymbol y_\ell^{(m)}(\mathcal{N}_i) \equiv \cup_{k=1}^{\ell-1} \{ Y^{(m)}(\boldsymbol s):\boldsymbol s \in \mathcal{N}_i \cap \mathcal{C}_k\}\, \bigcup \,\cup_{k=\ell+1}^{Q} \{ Y^{(m-1)}(\boldsymbol s):\boldsymbol s \in \mathcal{N}_i \cap \mathcal{C}_k \},
    \]
    where the second set union is treated as empty if \(\ell=Q\).
  \item
    For \(Q>2\), repeat step 3 for each \(\ell=3,\ldots,Q\).
  \end{enumerate}
\end{enumerate}
In each Gibbs iteration above, observations with locations in the \(\ell\)th conclique \(\mathcal{C}_\ell\) are updated conditionally on observations associated with other concliques, with observations from concliques \(\mathcal{C}_1,\ldots,\mathcal{C}_{\ell-1}\) being updated before conclique \(\mathcal{C}_{\ell}\). Crucially, at each conclique update, neighboring observations needed for defining a conditional distribution do not, by definition, belong to the conclique being updated. Additionally, note that any valid simulation approach may be used for an entire conclique update under independence (e.g., direct or acceptance sampling in Steps 1 or 3). The supplement describes two other possible conclique-based Gibbs samplers involving randomization in the order of conclique updates.

\hypertarget{theoretical-properties}{%
\subsection{Theoretical properties}\label{theoretical-properties}}

We mention two theoretical aspects about the conclique-based Gibbs sampler (CGS) regarding its validity and ergodicity. Let \(\underline{F}\) denote the joint distribution for \((Y(\boldsymbol s_1), \dots, Y(\boldsymbol s_n))\), that corresponds to the full conditionals (\ref{eqn:1}) for the MRF model. Firstly, under a mild support condition in Theorem \ref{thm:gibbs_harris} (that also holds under the typical positivity condition assumed in MRF formulations (c.f. Besag \protect\hyperlink{ref-besag1974spatial}{1974})), the CGS is guaranteed to capture \(\underline{F}\) as the number of Gibbs iterations increase and, hence, the sampler is Harris ergodic (cf.~Athreya and Lahiri \protect\hyperlink{ref-athreya2006measure}{2006}). To state the result, let \(\mathcal{X} \subset \mathbb{R}^n\) denote the support of the joint data distribution \(\underline{F}\) (e.g., with respect to a density/mass function) and let \(P^{(m)}(x, A)\), \(A\in\mathcal{F}\), denote the transition distribution of the CGS after \(m \geq 1\) complete iterations from an initializing point \(x \in \mathcal{X}\),
where \(\mathcal{F}\) represents a \(\sigma\)-algebra associated with \(\mathcal{X}\subset \mathbb{R}^n\).
\begin{theorem}
\label{thm:gibbs_harris}
Suppose that $\mathcal{X} = \mathcal{X}_1 \times \cdots \times \mathcal{X}_Q$ holds, where $\mathcal{X}_\ell$ denotes the marginal support of observations $\{Y(s_i): s_i \in \mathcal{C}_\ell\}$ with locations in conclique $\mathcal{C}_\ell$, $\ell=1, \dots, Q$. Then, the CGS is Harris ergodic with stationary distribution $\underline{F}(\cdot)$ and, for any initialization $x\in\mathcal{X}$, the sampler converges monotonically in total variation as the number $m$ of iterations increase, i.e.,
\begin{equation}
\label{eqn:CGS}
\sup_{A \in \mathcal{F}}| P^{(m)}(x,A) -\underline{F}(A)| \downarrow 0\quad \text{ as } m \to \infty.
\end{equation}
\end{theorem}
Additionally, for a general class of MRF specifications exhibiting two concliques, the CGS is also provably geometrically ergodic or, equivalently, exhibits a geometrically fast mixing rate as a function of the number \(m \geq 1\) of iterations: it holds in (\ref{eqn:CGS}) that \(\sup_{A \in \mathcal{F}}\vert P^{(m)}(x, A) - \underline{F}(A) \vert \le G(x)t^m\) for any \(x \in \mathcal{X}\) and for some real-valued function \(G: \mathcal{X} \rightarrow \mathbb{R}\) and constant \(t \in (0,1)\). MRF models with two concliques, while specialized, are often encountered in applications (e.g., four-nearest neighborhoods in spatial modeling (Figure \ref{fig:concliques}) or the network example of Section \ref{simulation-of-a-large-network}). In contrast, a similar result with the single-site sequential Gibbs sampler is not possible to establish for these MRF models, or more generally, as theory for geometric ergodicity of Gibbs samplers is essentially restricted to two-component Gibbs; see Johnson and Burbank (\protect\hyperlink{ref-johnson2015geometric}{2015}) and references therein. In this sense, the CGS allows additional convergence properties to be shown which are theoretically intractable with standard Gibbs sampling. The supplement establishes the geometric ergodicity of the CGS for several types of conditional distributions (\ref{eqn:1}) for \(Y(\boldsymbol s_i)\), \(i=1,\ldots,n\), having \(Q = 2\) concliques and bounded support (e.g., autologistic, Beta, or windsorized Poisson distributions from Cressie (\protect\hyperlink{ref-cressie1993statistics}{1993}) and Kaiser and Cressie (\protect\hyperlink{ref-kaiser1997modeling}{1997})) as well as for conditional gamma, inverse Gaussian, and Gaussian models with four-nearest neighborhoods. For reference next, the latter has a conditional density (with parameters \(|\eta| < 0.25\) and \(\alpha\)) as
\begin{equation}
\label{eqn:3}
f_i(y(\boldsymbol s_i)|\boldsymbol y(\mathcal{N}_i)) = \frac{1}{\sqrt{2\pi}\tau}\exp\left\{-\frac{1}{2\tau^2}(y(\boldsymbol s_i) - \mu(\boldsymbol s_i))^2\right\}, \quad y(\boldsymbol s_i) \in \mathbb{R},
\end{equation}
involving a conditional variance \(\tau^2\) and conditional mean
\(\mu(\boldsymbol s_i) = \alpha + \eta\sum_{s_j \in \mathcal{N}_i}\{y(s_j) - \alpha\}\).

\hypertarget{computational-speed}{%
\subsection{Computational speed}\label{computational-speed}}

For repeated simulation for MRF models, the conclique-based Gibbs sampler (CGS) is again intended to be computationally faster than standard single-site Gibbs sampling. As initial illustration, we evaluated timing results for both samplers in generating data from the conditional Gaussian specification (\ref{eqn:3}) (i.e., four-nearest neighbors) with various spatial grid sizes \(n = m \times m\) for \(m = 5, 10, 20, 30, 50, 75\), as well as numbers \(M\) of sampling iterations for \(M=100, 1000, 5000, 10000\). We chose \(\alpha=0\), \(\tau^2=1\) and \(\eta=0.2\) in (\ref{eqn:3}), though the exact values are immaterial to the timing study, and we implemented both samplers using C++ implementations in an available R package \texttt{conclique} (Kaplan \protect\hyperlink{ref-conclique}{2019}) on a 1.7 GHz processor. To facilitate comparison, the timing results do not include initial computational overheads for the methods that need to be performed once per sample size \(n\) regardless of \(M\). (Both samplers require listing neighbors for each observation under the model and the CGS additionally requires concliques. As reference, if we consider implementations based purely in \texttt{R}, then about \(0.25\) seconds are required to enumerate neighbors for the largest grid \(n=75\times 75\) and, for finding \(Q=2\) concliques, the times required for sample sizes \(n=20\times 20\), \(50\times 50\) and \(75 \times 75\) are respectively: \(\approx (0,0,0)\) seconds based on the geometrical pattern in Figure \ref{fig:concliques} and \(\approx (0.05, 1, 4)\) or \(\approx (0.3, 8, 40)\) seconds based on Welch-Powell or DSatur algorithms from Section \ref{finding-concliques}; the Welch-Powell search is faster but is sensitive to observation labeling and can yield 4 concliques here.)
\begin{figure}
\centering
\includegraphics{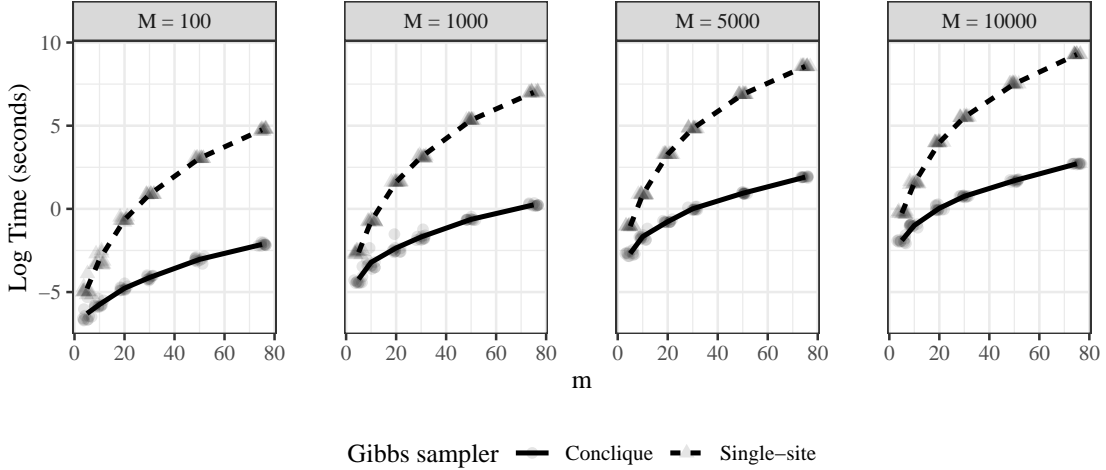}
\caption{\label{fig:timings-plot}\label{fig:timings}Log-times for simulation of \(M=100, 1000, 5000, 10000\) four-nearest neighbor MRF datasets on a lattice of size \(m \times m\) for various \(m = 5, 10, 20, 30, 50, 75\), using standard single-site and conclique-based Gibbs samplers (repeated 10 times (dots) with mean log run times as lines).}
\end{figure}
\par

Figure \ref{fig:timings} summarizes log running times for simulating \(M\) data sets from the Gaussian MRF model on a grid size \(n = m\times m\) for various \(m\) and \(M\). While the time difference between samplers is minimal over small grids (e.g., \(5 \times 5\)), the time savings with the CGS is substantial as grid size increases. For example, to simulate \(10,000\) spatial data sets of size \(75 \times 75\), the CGS required \(15.05\) seconds compared to
\(\ensuremath{1.076\times 10^{4}}\) seconds (\(\approx 2.99\) hours) with the standard Gibbs sampler. While computational time is linear in the number \(M\) of iterations with both samplers, computational time grows exponentially larger for the standard Gibbs sampler compared to the CGS as spatial sample size \(n = m\times m\) increases through \(m\). By its small number \(Q=2\) of Gibbs steps/concliques, the CGS is dramatically more time-efficient here for simulating large collections of even moderately sized samples.

\hypertarget{numerical-comparisons-motivated-by-spatial-bootstrap}{%
\section{Numerical comparisons motivated by spatial bootstrap}\label{numerical-comparisons-motivated-by-spatial-bootstrap}}

Section \ref{a-numerical-study-of-simulation-efficacy} summarizes a simulation study to compare the conclique-based Gibbs sampler (CGS) to the standard single-site Gibbs sampler in terms of mixing and computing costs. The simulation design involves three MRF models of increasing complexity for spatial binary data, which are motivated from a bootstrap application for modeling the presence (1) or
absence (0) of footrot in endive plants on a grid (Besag \protect\hyperlink{ref-besag1977some}{1977}). Section \ref{background-binary-models-spatial-bootstrap} provides brief background on these models, while the supplement details the spatial bootstrap application.

\hypertarget{background-binary-models-spatial-bootstrap}{%
\subsection{Background: Binary models \& spatial bootstrap}\label{background-binary-models-spatial-bootstrap}}

For the endive data, three centered autologistic models were considered as: (a) isotropic (Besag \protect\hyperlink{ref-besag1977some}{1977}; Caragea and Kaiser \protect\hyperlink{ref-caragea2009autologistic}{2009}), (b) ansiotropic with two dependence parameters, or (c) as in (b) but with large scale structure determined by regression on the horizontal coordinate \(u_i\) of each spatial location \(\boldsymbol s_i=(u_i,v_i)\). Each model has a resulting conditional mass function

\[
f_i(y(\boldsymbol {s}_i)|\boldsymbol {y}(\mathcal{N}_i)) = \frac{\exp[y(\boldsymbol {s}_i)A_i\big\{\boldsymbol {y}(\mathcal{N}_i)\big\} ]}
{1 +\exp[y(\boldsymbol {s}_i)A_i\big\{\boldsymbol {y}(\mathcal{N}_i)\big\} ]}, \quad y(\boldsymbol {s}_i) = 0, 1,
\]

of the form (\ref{eqn:1})-(\ref{eqn:2}) with natural parameter functions, \(A_i\big\{\boldsymbol {y}(\mathcal{N}_i)\big\}\), given in Table \ref{tab:natural-params} involving a four-nearest neighborhood and parameters \((\kappa, \eta)\) for Model (a), \((\kappa, \eta_u,\eta_v)\) for Model (b), and \((\beta_0,\beta_1, \eta_u,\eta_v)\) for Model (c).
\begin{table}[t]
\caption{Full conditional distributions (centered autologistic) of three binary MRF models. Fitted values were $(\hat{\eta},\hat{\kappa}) = (0.816,0.123)$, $(\hat{\eta}_u, \hat{\eta}_v, \hat{\kappa})=(0.958,0.660, 0.125)$ and $(\hat{\eta}_u, \hat{\eta}_v, \hat{\beta}_0, \hat{\beta}_1)=(0.000,0.004, -1.600, -0.001)$ under Models (a)-(c), respectively.}
\label{tab:natural-params}
\centering
\begin{tabular}{|p{.1in}  p{6.0in} |}
\hline
(a)& Isotropic with $A_i\{\boldsymbol y(\mathcal{N}_i)\} = \log\left(\frac{\kappa}{1-\kappa}\right) + \eta\sum_{\boldsymbol s_j \in \mathcal{N}_i}\{y(\boldsymbol s_j) - \kappa\}$, $\kappa\in(0,1)$, $\eta\in\mathbb{R}$, \& $\mathcal{N}_i =\{\boldsymbol s_i\pm (1,0),\boldsymbol s_i\pm (0,1)\}$     \\[.3cm]
(b) &Ansiotropic with $A_i\{\boldsymbol y(\mathcal{N}_i)\} = \log\left(\frac{\kappa}{1-\kappa}\right) + \eta_u\sum_{\boldsymbol s_j \in N_{u,i}}\{y(\boldsymbol s_j) - \kappa\} + \eta_v\sum_{\boldsymbol s_j \in N_{v,i}}\{y(\boldsymbol s_j) - \kappa\}$, $\kappa \in (0,1)$, horizontal/vertical dependence $\eta_u,\eta_v\in\mathbb{R}$, \& neighbors $\mathcal{N}_{u,i}=\{\boldsymbol s_i \pm (1,0)\}$, $\mathcal{N}_{v,i}=\{\boldsymbol s_i \pm (0,1)\}$   \\[.3cm]
(c) & like (b) with $A_i\{\boldsymbol y(\mathcal{N}_i)\}  = \log\left(\frac{\kappa_i}{1-\kappa_i}\right) + \eta_u\sum_{\boldsymbol s_j \in N_{u,i}}\{y(\boldsymbol s_j) - \kappa_i\} + \eta_v\sum_{\boldsymbol s_j \in N_{v,i}}\{y(\boldsymbol s_j) - \kappa_i\}$
but with $\kappa_i$ determined by  logistic regression $\mathrm{logit}(\kappa_i)  = \beta_0 + \beta_1 u_i$
on horizontal coordinate $u_i$ of location $\boldsymbol s_i=(u_i,v_i)$,  \& $\beta_0,\beta_1\in\mathbb{R}$    \\
\hline
\end{tabular}
\end{table}
As described in the supplement, parameter estimates were obtained by pseudo-likelihood and, based on these, both the CGS and standard Gibbs samplers were applied to generate \(10,000\) spatial datasets under each model in order to obtain reference distributions for goodness-of-fit statistics and confidence intervals. This represents a parametric bootstrap approximation, where simulation speed and efficiency are useful in rendering a large number of spatial data sets from differing models. Both samplers produced nearly identitical bootstrap approximations, though the standard Gibbs approach required tens of minutes for each model compared to tens of seconds for the CGS. These findings suggest, though, that both samplers exhibit similar algorithmic, or mixing, efficiency which we next examine further.

\hypertarget{a-numerical-study-of-simulation-efficacy}{%
\subsection{A numerical study of simulation efficacy}\label{a-numerical-study-of-simulation-efficacy}}

Here we numerically compare the conclique-based Gibbs sampler (CGS) to the standard single-site Gibbs sampler, as well as the Swendsen-Wang algorithm (an auxiliary variable method from Sec.\secspace\ref{other-simulation-approaches}) in terms of mixing effectiveness (or algorithmic capability to produce approximately independent samples from the target joint data distribution) in addition to computational speed (or timing demands). These contributors to Markov chain Monte Carlo (MCMC) efficiency are quantified using measures from Turek et al. (\protect\hyperlink{ref-turek2017automated}{2017}). We compare the samplers for simulating spatial data \(\{Y(\boldsymbol s_i)\}_{i=1}^n\) on a \(40\times 40\) grid from the three binary MRF models of varying complexity, namely, Models (a)-(c) as prescribed by fitted parameter values in Table \ref{tab:natural-params}, as well as a fourth binary MRF model, Model (d), that is close to criticality (\(\eta = 0.88\)) in its parameterization with no large-scale mean parameter (i.e., its analog in Table \ref{tab:natural-params} would be \(A_i(\boldsymbol {y}(\mathcal{N}_i)) \equiv \eta \sum_{\boldsymbol {s}_j \in \mathcal{N}_i} y(\boldsymbol {s}_j)\)).

To assess mixing or algorithmic efficiency, we consider a quantity
\[
Alg = \min\limits_{1 \le i \le n}\left\{\left( 1 + 2\sum\limits_{j = 1}^\infty \rho_i(j)\right)^{-1}\right\},
\]
corresponding to the location-wise minimum of inverse integrated autocorrelations (cf.~Roberts and Rosenthal \protect\hyperlink{ref-roberts2001optimal}{2001}; Turek et al. \protect\hyperlink{ref-turek2017automated}{2017}), where \(\rho_i(j)\), \(j \geq 1\) denotes the autocorrelation function for the chain generations of observation \(Y(\boldsymbol s_i)\), \(i=1,\ldots,n\equiv 40\times 40\). For \(M\) full iterations of a Gibbs sampler, the value \(Alg\cdot M\) approximates the number of essentially independent data sets, after adjusting for the largest autocorrelation among MCMC iterations at a sampling location. Small values of \(Alg\) then indicate poor mixing properties for a sampler (typically \(Alg \leq 1\) with \(Alg\) tending to zero under increasing positive dependence in MCMC iterations). For each MRF Model (a)-(d) and type of sampler here, we obtained an initial estimate of \(Alg\) from a kernel estimator applied to the sample autocorrelations from a chain realization, as given in the package \texttt{LaplacesDemon} (Statisticat, LLC. \protect\hyperlink{ref-laplacesdemon}{2016}). Table \ref{tab:eff-results} reports final values of \(Alg\) determined by the average of such estimations from \(10\) chains, each with different starting values and \(10,000\) iterations. In addition to this algorithmic efficiency \(Alg\) in Table \ref{tab:eff-results}, we also provide computational cost for all four MRF models as measured in run-time per MCMC iteration (Turek et al. \protect\hyperlink{ref-turek2017automated}{2017}). This quantity, denoted as \(Comp\) in Table \ref{tab:eff-results}, represents the computing time (in seconds) for one complete data generation under each sampler (again using C++ implementations), without overhead due to intialization; Sec.\secspace \ref{computational-speed} describes the latter, which remains relevant here. Reported values of \(Comp\) are averages from \(20,000\) time recordings.

\par
\begin{table}[ht]
\centering
\begin{tabular}{|l|rr|rr|rr|rr|}
\hline
Algorithm & \multicolumn{2}{|c|}{Model (a)} & \multicolumn{2}{|c|}{Model (b)} & \multicolumn{2}{|c|}{Model (c)} & \multicolumn{2}{|c|}{Model (d)}\\
\cline{2-9}
 & $Alg$ & $Comp$ & $Alg$ & $Comp$ & $Alg$ & $Comp$ & $Alg$ & $Comp$ \\
\hline
Conclique Gibbs & $0.805$ & $\ensuremath{3.1\times 10^{-4}}$ & $0.747$ & $\ensuremath{3.5\times 10^{-4}}$ & $0.7$ & $\ensuremath{3.6\times 10^{-4}}$  & $0.715$ & $\ensuremath{2.5\times 10^{-4}}$\\
Standard Gibbs & $0.806$ & $0.023$ & $0.751$ & $0.027$ & $0.714$ & $0.02$ & $0.71$ & $0.02$\\
Swendsen-Wang & $0.863$ & $0.036$ & $0.869$ & $0.058$ & $0.515$ & $0.055$ & $0.863$ & $0.047$\\
\hline
\end{tabular}
\caption{Algorithmic $Alg$ and computational $Comp$ measures of simulation efficiency for four autologistic models on a $40 \times 40$ grid. Large $Alg$ and small $Comp$ values are preferable. }
\label{tab:eff-results}
\end{table}
Both conclique-based and single-site Gibbs samplers have similar algorithmic/mixing efficiencies in Table \ref{tab:eff-results}, agreeing with other contexts encountered in our investigations. To obtain an effective number of draws from a joint data distribution, both samplers generally require more iterations as the underlying model becomes more complex (Models (a) through (c)) or exhibits extreme dependence (Model (d)), though the per-iteration time-cost of each sampler remains fairly unchanged for these models. Additionally, Table \ref{tab:eff-results} supports the findings of Higdon (\protect\hyperlink{ref-higdon1994spatial}{1994}) and Hurn (\protect\hyperlink{ref-hurn1997difficulties}{1997}) that the Swendsen-Wang algorithm improves mixing over the standard Gibbs sampler when there is a constant global mean structure (Models (a)-(b)) or when the dependence is very high (Model (d)), but experiences a slowing down when there is non-constant large-scale structure (Model (c)). For all cases though, the conclique-based sampler is at least 55 times faster than the standard sampler and at least at least 117 times faster than the Swendsen-Wang algorithm. As a result, if mixing \(Alg\) and timing \(Comp\) rates from Table \ref{tab:eff-results} are counterbalanced into one overall cost \(Comp/Alg\), representing the computing time required to draw one effectively-independent data sample, then the conclique-based sampler continues to exhibit benefits, being at least 54 (or 116) times faster than standard Gibbs (or Swendsen-Wang). Consequently, the actual time savings of the conclique-based sampler over the single-site Gibbs and the Swendsen-Wang algorithm can be quite substantial, particularly as the desired number of MCMC iterations grows.

\hypertarget{illustrations-with-networks}{%
\section{Illustrations with networks}\label{illustrations-with-networks}}

The conclique-based Gibbs sampler (CGS) may also apply to simulation from MRF specifications of random networks and graphs. Section \ref{exponential-random-graphs-with-incidence-neighbors} provides an illustration with a class of exponential random graph models (cf.~Robins et al. \protect\hyperlink{ref-robins2007introduction}{2007}), while Section \ref{simulation-of-a-large-network} considers simulation of a large network using a local structure graph model (Casleton, Nordman, and Kaiser \protect\hyperlink{ref-casleton2017local}{2017}).

\hypertarget{exponential-random-graphs-with-incidence-neighbors}{%
\subsection{Exponential random graphs with incidence neighbors}\label{exponential-random-graphs-with-incidence-neighbors}}

Suppose random variables \(Y(\boldsymbol{s}_i) \in \{0,1\}\) denote the presence/absence of random edges at the \(n \equiv {V \choose 2}\) edge locations \(\{\boldsymbol{s}_1,\dots,\boldsymbol{s}_n\}\) (i.e., each \(\boldsymbol{s}_i=\{v_{i1},v_{i2}\}\) marking two vertices \(v_{i1},v_{i2}\)) in a simple undirected graph with \(V\) vertices, and consider a MRF model for graph edges defined through binary conditional densities (\ref{eqn:1}) having ``incidence'' neighborhoods \(\mathcal{N}_i\) from Section \ref{concliques} (i.e., neighbors \(\boldsymbol{s}_i\), \(\boldsymbol{s}_j\) share some vertex). Such conditional densities and neighborhoods are induced, for example, by \emph{joint distributions} for random graphs having an exponential form based on counts of triangles and \(k\)-stars (or linear combinations of these); see Frank and Strauss (\protect\hyperlink{ref-frank1986markov}{1986}) (e.g., homogeneous Markov graphs), Wasserman and Pattison (\protect\hyperlink{ref-wasserman1996logit}{1996}) and Snijders et al. (\protect\hyperlink{ref-snijders2006new}{2006}) (e.g., alterating \(k\)-stars) for examples, representing special types of exponential random graph models. To illustrate, for each edge location \(\boldsymbol{s}_i\), define neighborhood statistics \(t_{i}^*\equiv 2^{-1} \sum_{\boldsymbol{s}_j \in \mathcal{N}_i} y(\boldsymbol{s}_j)\) and
\(t_{i}^\blacktriangle\equiv 6^{-1} \sum_{(\boldsymbol{s}_j,\boldsymbol{s}_k) \in \mathcal{T}_i} y(\boldsymbol{s}_j)y(\boldsymbol{s}_k)\) for a set \(\mathcal{T}_i \equiv \{(\boldsymbol{s}_j,\boldsymbol{s}_k) : \boldsymbol{s}_i,\boldsymbol{s}_j,\boldsymbol{s}_k \text{ are neighbors}, i \neq j \neq k \}\).\\
Then, the count statistics \(\sum_{i=1}^n y(\boldsymbol{s}_i)\),
\(\sum_{i=1}^n y(\boldsymbol{s}_i) \cdot t_{i}^*\), \(\sum_{i=1}^n y(\boldsymbol{s}_i) \cdot t_{i}^\blacktriangle\) represent numbers of 1-stars, 2-stars and triangles, respectively, and the so-called ``triad'' graph model (Frank and Strauss \protect\hyperlink{ref-frank1986markov}{1986}) has a joint exponential distribution
\begin{equation}
\label{eqn:triad}
P(Y(\boldsymbol{s}_1)= y(\boldsymbol{s}_1),\ldots,Y(\boldsymbol{s}_n)= y(\boldsymbol{s}_n))\propto \exp\left[  \sum_{i=1}^n y(\boldsymbol{s}_i) \left(\rho +  \sigma t_{i}^* + \tau t_i^\blacktriangle  \right) \right],
\end{equation}
expressed as a linear combination of these counts in parameters \((\rho,\sigma,\tau)\).
As our notation suggests, the triad joint (\ref{eqn:triad}) consequently induces conditional densities \(P(Y(\boldsymbol{s}_i)= y(\boldsymbol{s}_i)| y(\boldsymbol{s}_j), j\neq i) \propto \exp\left[ y(\boldsymbol{s}_i) (\rho + 2\sigma t_{i}^* + 3\tau t_{i}^\blacktriangle) \right]\), \(y(\boldsymbol{s}_i)\in\{0,1\}\),
of form (\ref{eqn:1})-(\ref{eqn:2}) with ``incidence'' neighbors. To simulate from such random graph models, single-site Gibbs sampling based on full conditional densities is a standard approach (cf.~Snijders et al. \protect\hyperlink{ref-snijders2006new}{2006}), though we next explain that the CGS can offer major speed improvements.

Under the triad model (\ref{eqn:triad}) or any other graph model inducing incidence neighborhoods, the same concliques apply. Note that the basic bound in (\ref{eqn:bound}) (cf.~Sec \ref{finding-concliques}) gives that, for any number \(V>2\) of vertices, the number \(Q\) of concliques required under incidence neighborhoods cannot be more than \(2V - 3\), due to neighborhood sizes \(|\mathcal{N}_i| = 2(V - 2)\). Hence, the conclique number \(Q\) needed is smaller order compared to the total number \(n = V (V - 1)/2\) of edge variables \(Y(\boldsymbol s_i)\), which implies that the CGS can have large speed advantages over the standard single-site Gibbs sampler with these graph models. While concliques may be found by graph coloring algorithms (cf.~Welch-Powell of Section \ref{finding-concliques}), a minimal covering of \(Q=2 \lceil V/2 \rceil - 1\) concliques also exists based on a concrete geometrical description developed in the supplement; the DSatur coloring applied in \texttt{R} also finds these minimal concliques, requiring about \(2\) or \(30\) seconds with \(V = 50\) or \(100\) vertices. We use \(Q=2\lceil V/2 \rceil -1\) concliques, though timing results to follow would remain qualitatively similar for any conclique number with \(Q\leq 2V\ll n\). For illustration of the CGS, we simulate data \(\{Y(\boldsymbol {s}_1),\ldots,Y(\boldsymbol {s}_n)\}\) from a conditional model (\ref{eqn:1})-(\ref{eqn:2}) with incidence neighborhoods \(\mathcal{N}_i\), where \(Y(\boldsymbol s_i)\) given its neighbors \(\boldsymbol y(\mathcal{N}_i)\) is Bernoulli(\(p(\boldsymbol s_i, \kappa, \eta)\)) with
\[
\text{logit}(p(\boldsymbol s_i, \kappa, \eta)) = \text{logit}(\kappa) + \frac{\eta_1}{|\mathcal{N}_i|}\sum\limits_{\boldsymbol s_j \in \mathcal{N}_i}\{y(\boldsymbol s_j) - \kappa\} +\frac{\eta_2}{|\mathcal{T}_{i}|}\sum_{\{\boldsymbol{s}_j,\boldsymbol{s}_k\}\in\mathcal{T}_i} \{y(\boldsymbol{s}_j)y(\boldsymbol{s}_k) - \kappa^2\},
\]
involving scale \(\kappa \in (0,1)\) and dependence \(\eta_1, \eta_2 \in \mathbb{R}\) parameters along with sizes \(|\mathcal{N}_i|=2(V-2)=|\mathcal{T}_i|\) in a centered parameterization of the triad model to facilitate interpretation of parameters (cf.~Casleton, Nordman, and Kaiser \protect\hyperlink{ref-casleton2017local}{2017}).
\begin{figure}
\centering
\includegraphics{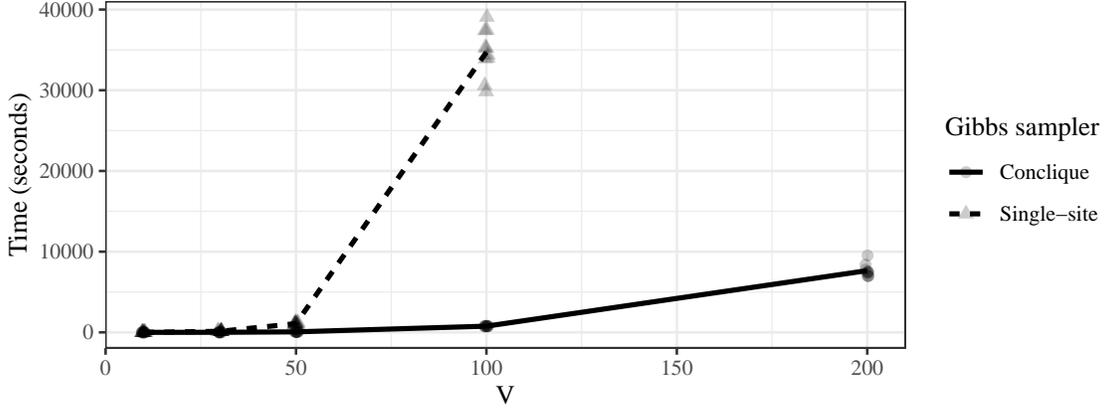}
\caption{\label{fig:timings2}\label{fig:graph-timing}Simulation time for \(1,000\) iterations under both standard single-site and conclique-based Gibbs samplers (CGS), in generating networks with \(V = 10, 30, 50, 100\) vertices. Each simulation was run 10 times (transparent dots) and the mean run time is shown via lines and opaque dots. The standard sampler becomes infeasible when \(V >100\) (e.g., requiring more than 2.5 days when \(V=200\) compared to 2.1 hours (about 7600 seconds) for CGS).}
\end{figure}
For various numbers \(V\) of vertices, Figure \ref{fig:graph-timing} shows the computational times needed for \(1,000\) data generations of size \(n = {V \choose 2}\) graphs with both samplers (for parameters \(\kappa=0.2\), \(\eta_1 = \eta_2 = 0.5\) in the R package \texttt{conclique} (Kaplan \protect\hyperlink{ref-conclique}{2019}) on a 1.7 GHz processor). The computational speed of the CGS is superior, growing about linearly with the number \(V \approx \sqrt{2n}\) of vertices. With \(V = 100\) vertices, for example, \(1,000\) samples required \(765.51\) seconds (\(12.76\) minutes) with the CGS and \(\ensuremath{3.472\times 10^{4}}\) seconds (\(9.64\) hours) with the standard sampler. The latter becomes impractical when \(V > 100\), which also agrees with studies in Schweinberger and Handcock (\protect\hyperlink{ref-schweinberger2015local}{2015}) where standard Gibbs simulation of graphs with \(V = 100\) vertices were considered nearly infeasible under the triad model.

For the incidence neighbor-type of exponential graph models here, neither the single-site Gibbs sampler nor the CGS is computationally satisfactory for graphs with vertices \(V>1000\), and mixing efficiency can depend heavily on model parameters (cf.~Handcock et al. \protect\hyperlink{ref-handcock2003assessing}{2003}), though the form of parameterization may help (cf.~Snijders et al. \protect\hyperlink{ref-snijders2006new}{2006}; Casleton, Nordman, and Kaiser \protect\hyperlink{ref-casleton2017local}{2017}). Still, single-site sampling often plays a basic role in simulating from such models (e.g., underlying ``perfect sampling'' algorithms (Butts \protect\hyperlink{ref-butts2018perfect}{2018}) and other software (Hunter et al. \protect\hyperlink{ref-hunter2008ergm}{2008})), and the CGS can offer better computational scalability. Simulation of such graphs with vertices \(V\) in the hundreds (e.g., some organizational networks) is then practical with the CGS in a way that has not been previously possible with standard Gibbs sampling.

\hypertarget{simulation-of-a-large-network}{%
\subsection{Simulation of a large network}\label{simulation-of-a-large-network}}

Chyzh and Kaiser (\protect\hyperlink{ref-chyzh2019local}{2019}) applied a network model to data on the formation of defense alliances between countries from 1946 to 2007. Nodes in the network consisted of country-year combinations determined to be ``politically relevant'' for alliances, with edges between node pairs per year indicating existence of a defense agreement. Thus, the same two countries in different years would define two different potential edges. There were a total of \(n=45,513\) possible edges in the application. Chyzh and Kaiser (\protect\hyperlink{ref-chyzh2019local}{2019}) used a local structure graph model (Casleton, Nordman, and Kaiser \protect\hyperlink{ref-casleton2017local}{2017}), which formulates probabilities of edge realization in a binary MRF specification for random variables with \(Y_i\equiv Y(\boldsymbol {s}_i)=1\) if an edge \(i\) is present and \(0\) otherwise, \(i=1, \ldots, n\). Specifically, the full conditional probability mass functions (\ref{eqn:1})-(\ref{eqn:2}) were
\begin{equation} \label{CKmodel}
\mathrm{logit} [P(Y_i=1| \{y_j: \, j \neq i\})]  = \log\left( \frac{\kappa_i}{1-\kappa_i} \right) + \eta \sum_{j =1}^{m} d_{i,j} \left( y_j - \kappa_j \right),
\end{equation}
where \(d_{i,j}\) is a political-ideological distance between edges \(i,j\) (two pairs of countries) constructed from annual voting records in the United Nations; see Chyzh and Kaiser (\protect\hyperlink{ref-chyzh2019local}{2019}) for details.
In the application, the values of \(\kappa_i\) were further modeled as,
\[
\log\left(\frac{\kappa_i}{1-\kappa_i} \right) =  \beta_0 + \beta_1 x_{1,i} + \beta_2 x_{2,i} + \beta_3 x_{3,i},
\]
where \(x_{1,i}\) was a variable giving the ratio of sizes of militaries for the two countries that define potential edge \(i\), \(x_{2,i}\) was log-transformed total trade between the two countries, and \(x_{3,i}\) was an indicator of joint democracy, having a value of \(1\) if both countries had democratically elected governments and \(0\) otherwise. Here \(\mbbeta=(\beta_0, \beta_1, \beta_2, \beta_3)\) controls large-scale structure of the network, while \(\eta\) controls small-scale structure. Using a composite likelihood in the form of Besag's original pseudo-likelihood (Besag \protect\hyperlink{ref-besag1975statistical}{1975}), Chyzh and Kaiser (\protect\hyperlink{ref-chyzh2019local}{2019}) obtained estimates \(\hat{\beta}_0 = 0.094\), \(\hat{\beta}_1=-2.363\), \(\hat{\beta}_2 = 0.015\), \(\hat{\beta}_3 = 0.884\), and \(\hat{\eta}=0.016\). The positive value for \(\hat{\eta}\) indicated that large ideological distance between two pairs of countries (or potential edges \(i,j\)) influence the formation of defense alliances (i.e., \(Y_i=1\) and \(Y_j=1\)).

Exploratory analysis of these data suggests that the relation between dependence and ideological distance between edges may be driven by a relatively small set of edge pairs that are separated by fairly large distances. To examine this possibility, we conducted a Monte Carlo goodness-of-fit procedure by replacing \(d_{i,j}\) in (\ref{CKmodel}) by \(I(d_{i,j} \geq 3)\), where \(I(A)\) denotes an indicator function having a value of \(1\) if event \(A\) is true and a value of \(0\) otherwise. Note that this is the same as defining neighborhoods as edges that are separated by distances of at least \(3\). We then simulated graphs using the estimates of Chyzh and Kaiser (\protect\hyperlink{ref-chyzh2019local}{2019}) under the conjecture that only edges separated by large distances influenced the results that those authors report. Due to the number of graph edges, we applied the conclique-based sampler (CGS) to facilitate timely simulation. Concliques were found by the DSatur graph coloring algorithm (cf.~Sec.\secspace\ref{finding-concliques}). While the largest neighborhood \(\mathcal{N}_i =\{j : d_{i,j} \geq 3 \}\) had size \(554\) under the hypothesized model (so that the upper bound \(555\) from (\ref{eqn:bound}) on the number \(Q\) of concliques required is already small compared to \(n = 45,513\)), the algothrim found \(Q=2\) concliques for partitioning the \(45,513\) possible edges.

The test statistic chosen was the proportion of pairs of edges with \(d_{i,j} \geq 3\) that
both assumed a value of 1. With \(T_m\) denoting this statistic for simulated graph
\(m = 1, \ldots, M\) and \(T_A\) denoting the value for the actual graph, a Monte Carlo p-value is \(p_M = M^{-1}\sum_{m=1}^M I(T_m \geq T_A)\), where \(I(\cdot)\) is again the indicator function. Either extreme small or extreme large values of \(p_M\) indicate disagreement of the data with the generating model. Using the data of Chyzh and Kaiser (\protect\hyperlink{ref-chyzh2019local}{2019}) available from the Political Analysis Dataverse (\url{https://dataverse.harvard.edu/dataverse/pan}),
the result was \(p_M = 0.0024\) based on \(M=2,500\) graph samples (after a burn-in of \(10,000\) iterations and thinning by \(100\), requiring about \(10\) minutes). Our conclusion, then, is that the suggestion arising from our exploratory analysis is not supported upon more formal assessment. It would appear that it is not only the pairs of potential edges separated by the largest distances that are driving the result that ideological distance between pairs of countries influences the formation of defense alliances.

\hypertarget{concluding-remarks}{%
\section{Concluding remarks}\label{concluding-remarks}}

Repeated simulation of data from MRF models is often important in statistical inference. Using concliques, we have presented a general Gibbs sampler for such simulation that can be much faster than the standard single-update Gibbs strategy. From a MRF model specification, the conclique-based Gibbs sampler (CGS) allows blocks of non-neighboring observations to be updated independently and simultaneously, where speed advantages were demonstrated in several numerical studies (Sec.\secspace\ref{conclique-based-gibbs-sampling}-\ref{illustrations-with-networks}). Large speed improvements over the sequential single-site sampler require the number \(Q\) of concliques to be effectively smaller than the sample size \(n\) to be simulated, though \(Q \leq n\) always holds and, in the worst case scenario (i.e., \(Q=n\) or all concliques of size 1), the CGS reduces to a standard single-site sampler. Hence, the CGS allows for faster iterations, though mixing efficiencies of both samplers appear comparable from numerical investigations. The CGS can work well in the same situations as the standard Gibbs sampler applies and, conversely, the CGS can exhibit slow mixing when the standard Gibbs sampler does as well (e.g., models near criticality from extreme dependence configurations, cf.~Sec.\secspace\ref{a-numerical-study-of-simulation-efficacy}).

Several areas of investigation exist with the CGS. For simulation of massive data sets, parallel computing appears possible with the CGS in a manner that would be unavailable in the standard Gibbs approach. In particular, the independent updates of all observations per each conclique are open to potential parallelization, which could induce further computational efficiencies. Another issue of research involves the determination of concliques from a MRF model specification. Identification of concliques shares connections to ``coloring problems'' in graph theory, and graph theoretic algorithms for vertex coloring may be applied to determine concliques (Sec.\secspace\ref{finding-concliques}). However, model information from the nature of neighborhoods themselves might also be investigated for prescribing concliques efficiently. Possibilities exist for the development of MRF models for spatial and network structures with a focus on the neighborhood ``geographies'' that might promote model simulation via concliques. Concliques might also be considered with other update strategies (e.g., Metropolis-Hastings within Gibbs) in the development of new samplers.

\section*{Supplemental Materials}

\textbf{Appendices}: Includes additional versions of the conclique Gibbs sampler, proofs of ergodicity results, a construction of concliques for graphs with incident neighborhoods, and a spatial bootstrap example. (Zip file)

\textbf{Code}: All code necessary to reproduce the results in this paper. (Zip file)

\textbf{Data}: An archive containing the data from Section \ref{simulation-of-a-large-network}. (Zip file)

\section*{Acknowledgements}

The authors are grateful to two reviewers, an Associate Editor and the Editor (Prof.~Tyler McCormick) for thoughtful comments and suggestions that greatly improved the manuscript. Research partially supported by NSF DMS-1310068, DMS-1613192, and DMS-1406747.

\hypertarget{references}{%
\section*{References}\label{references}}
\addcontentsline{toc}{section}{References}

\hypertarget{refs}{}
\leavevmode\hypertarget{ref-arnold2001conditionally}{}%
Arnold, Barry C, Enrique Castillo, and Jose Maria Sarabia. 2001. ``Conditionally Specified Distributions: An Introduction (with Comments and a Rejoinder by the Authors).'' \emph{Statistical Science} 16 (3): 249--74.

\leavevmode\hypertarget{ref-athreya2006measure}{}%
Athreya, Krishna B, and Soumendra N Lahiri. 2006. \emph{Measure Theory and Probability Theory}. Springer-Verlag New York.

\leavevmode\hypertarget{ref-besag1974spatial}{}%
Besag, Julian. 1974. ``Spatial Interaction and the Statistical Analysis of Lattice Systems.'' \emph{Journal of the Royal Statistical Society. Series B (Methodological)}, 192--236.

\leavevmode\hypertarget{ref-besag1975statistical}{}%
---------. 1975. ``Statistical Analysis of Non-Lattice Data.'' \emph{The Statistician}, 179--95.

\leavevmode\hypertarget{ref-besag1977some}{}%
---------. 1977. ``Some Methods of Statistical Analysis for Spatial Data.'' \emph{Bulletin of the International Statistical Institute} 47 (2): 77--92.

\leavevmode\hypertarget{ref-besag1993spatial}{}%
Besag, Julian, and Peter J Green. 1993. ``Spatial Statistics and Bayesian Computation.'' \emph{Journal of the Royal Statistical Society: Series B (Methodological)} 55 (1): 25--37.

\leavevmode\hypertarget{ref-besag1991bayesian}{}%
Besag, Julian, Jeremy York, and Annie Mollie. 1991. ``Bayesian Image Restoration with Two Applications in Spatial Statistics (with Discussion).'' \emph{Annals of the Institute of Statistical Mathematics} 43: 1--59.

\leavevmode\hypertarget{ref-brelaz1979new}{}%
Brélaz, Daniel. 1979. ``New Methods to Color the Vertices of a Graph.'' \emph{Communications of the ACM} 22 (4): 251--56.

\leavevmode\hypertarget{ref-butts2018perfect}{}%
Butts, Carter T. 2018. ``A Perfect Sampling Method for Exponential Family Random Graph Models.'' \emph{The Journal of Mathematical Sociology} 42 (1): 17--36.

\leavevmode\hypertarget{ref-caragea2009autologistic}{}%
Caragea, Petruţa C, and Mark S Kaiser. 2009. ``Autologistic Models with Interpretable Parameters.'' \emph{Journal of Agricultural, Biological, and Environmental Statistics} 14 (3): 281.

\leavevmode\hypertarget{ref-casleton2017local}{}%
Casleton, Emily, Daniel J Nordman, and Mark S Kaiser. 2017. ``A Local Structure Model for Network Analysis.'' \emph{Statistics and Its Interface} 10 (2): 355--67.

\leavevmode\hypertarget{ref-chyzh2019local}{}%
Chyzh, Olga V, and Mark S Kaiser. 2019. ``A Local Structure Graph Model: Modeling Formation of Network Edges as a Function of Other Edges.'' \emph{Political Analysis} In Press.

\leavevmode\hypertarget{ref-cressie1993statistics}{}%
Cressie, Noel. 1993. \emph{Statistics for Spatial Data: Wiley Series in Probability and Statistics}. Wiley: New York, NY, USA.

\leavevmode\hypertarget{ref-davies2013circulant}{}%
Davies, Tilman M, and David Bryant. 2013. ``On Circulant Embedding for Gaussian Random Fields in R.'' \emph{Journal of Statistical Software} 55 (9): 1--21.

\leavevmode\hypertarget{ref-frank1986markov}{}%
Frank, Ove, and David Strauss. 1986. ``Markov Graphs.'' \emph{Journal of the American Statistical Association} 81 (395): 832--42.

\leavevmode\hypertarget{ref-friel2004likelihood}{}%
Friel, Nial, and AN Pettitt. 2004. ``Likelihood Estimation and Inference for the Autologistic Model.'' \emph{Journal of Computational and Graphical Statistics} 13 (1): 232--46.

\leavevmode\hypertarget{ref-hammersley1971markov}{}%
Hammersley, John M, and Peter Clifford. 1971. ``Markov Fields on Finite Graphs and Lattices.'' \emph{Unpublished}.

\leavevmode\hypertarget{ref-handcock2003assessing}{}%
Handcock, Mark S, Garry Robins, Tom Snijders, Jim Moody, and Julian Besag. 2003. ``Assessing Degeneracy in Statistical Models of Social Networks.'' \url{http://www.csss.washington.edu/Papers}.

\leavevmode\hypertarget{ref-higdon1994spatial}{}%
Higdon, David M. 1994. ``Spatial Applications of Markov Chain Monte Carlo for Bayesian Inference.'' PhD thesis, Seattle: Department of Statistics, University of Washington.

\leavevmode\hypertarget{ref-higdon1998auxiliary}{}%
---------. 1998. ``Auxiliary Variable Methods for Markov Chain Monte Carlo with Applications.'' \emph{Journal of the American Statistical Association} 93 (442): 585--95.

\leavevmode\hypertarget{ref-hoff2002latent}{}%
Hoff, Peter D, Adrian E Raftery, and Mark S Handcock. 2002. ``Latent Space Approaches to Social Network Analysis.'' \emph{Journal of the American Statistical Association} 97 (460): 1090--8.

\leavevmode\hypertarget{ref-hughes2014ngspatial}{}%
Hughes, John. 2014. ``ngspatial: A Package for Fitting the Centered Autologistic and Sparse Spatial Generalized Linear Mixed Models for Areal Data.'' \emph{The R Journal} 6 (2): 81--95. \url{https://journal.r-project.org/archive/2014/RJ-2014-026/index.html}.

\leavevmode\hypertarget{ref-hughes2011autologistic}{}%
Hughes, John, Murali Haran, and Petruţa C Caragea. 2011. ``Autologistic Models for Binary Data on a Lattice.'' \emph{Environmetrics} 22 (7): 857--71.

\leavevmode\hypertarget{ref-hunter2008ergm}{}%
Hunter, David R, Mark S Handcock, Carter T Butts, Steven M Goodreau, and Martina Morris. 2008. ``Ergm: A Package to Fit, Simulate and Diagnose Exponential-Family Models for Networks.'' \emph{Journal of Statistical Software} 24 (3): nihpa54860.

\leavevmode\hypertarget{ref-mapcoloring}{}%
Hunziker, Philipp. 2017. \emph{MapColoring: Optimal Contrast Map Coloring}.

\leavevmode\hypertarget{ref-hurn1997difficulties}{}%
Hurn, Merrilee. 1997. ``Difficulties in the Use of Auxiliary Variables in Markov Chain Monte Carlo Methods.'' \emph{Statistics and Computing} 7 (1): 35--44.

\leavevmode\hypertarget{ref-husfeldt2015graph}{}%
Husfeldt, Thore. 2015. ``Graph Colouring Algorithms.'' In \emph{Topics in Chromatic Graph Theory}, edited by L. W. Beineke and Robin J. Wilson, 277--30. Encyclopedia of Mathematics and Its Applications. Cambridge: Cambridge University Press.

\leavevmode\hypertarget{ref-jensen2011graph}{}%
Jensen, Tommy R, and Bjarne Toft. 2011. \emph{Graph Coloring Problems}. Vol. 39. John Wiley \& Sons.

\leavevmode\hypertarget{ref-johnson2015geometric}{}%
Johnson, Alicia A, and Owen Burbank. 2015. ``Geometric Ergodicity and Scanning Strategies for Two-Component Gibbs Samplers.'' \emph{Communications in Statistics - Theory and Methods} 44 (15): 3125--45.

\leavevmode\hypertarget{ref-kaiser2009exploring}{}%
Kaiser, Mark S, and Petruţa C Caragea. 2009. ``Exploring Dependence with Data on Spatial Lattices.'' \emph{Biometrics} 65 (3): 857--65.

\leavevmode\hypertarget{ref-kaiser2012centered}{}%
Kaiser, Mark S, Petruţa C Caragea, and Kyoji Furukawa. 2012. ``Centered Parameterizations and Dependence Limitations in Markov Random Field Models.'' \emph{Journal of Statistical Planning and Inference} 142 (7): 1855--63.

\leavevmode\hypertarget{ref-kaiser1997modeling}{}%
Kaiser, Mark S, and Noel Cressie. 1997. ``Modeling Poisson Variables with Positive Spatial Dependence.'' \emph{Statistics \& Probability Letters} 35 (4): 423--32.

\leavevmode\hypertarget{ref-kaiser2000construction}{}%
---------. 2000. ``The Construction of Multivariate Distributions from Markov Random Fields.'' \emph{Journal of Multivariate Analysis} 73 (2): 199--220.

\leavevmode\hypertarget{ref-kaiser2012goodness}{}%
Kaiser, Mark S, Soumendra N Lahiri, and Daniel J Nordman. 2012. ``Goodness of Fit Tests for a Class of Markov Random Field Models.'' \emph{The Annals of Statistics} 40 (1): 104--30.

\leavevmode\hypertarget{ref-kaiser2012blockwise}{}%
Kaiser, Mark S, and Daniel J Nordman. 2012. ``Blockwise Empirical Likelihood for Spatial Markov Model Assessment.'' \emph{Statistics and Its Interface} 5 (3): 303--18.

\leavevmode\hypertarget{ref-conclique}{}%
Kaplan, Andee. 2019. \emph{Conclique: Gibbs Sampling for Spatial Data and Concliques}. \url{https://github.com/andeek/conclique}.

\leavevmode\hypertarget{ref-li2012markov}{}%
Li, Stan Z. 2012. \emph{Markov Random Field Modeling in Computer Vision}. Springer Science \& Business Media.

\leavevmode\hypertarget{ref-moller1999perfect}{}%
Møller, Jesper. 1999. ``Perfect Simulation of Conditionally Specified Models.'' \emph{Journal of the Royal Statistical Society: Series B (Statistical Methodology)} 61 (1): 251--64.

\leavevmode\hypertarget{ref-moller2003statistical}{}%
Møller, Jesper, and Rasmus Plenge Waagepetersen. 2003. \emph{Statistical Inference and Simulation for Spatial Point Processes}. CRC Press.

\leavevmode\hypertarget{ref-novikovpyclustering}{}%
Novikov, Andrei V. 2019. ``PyClustering: Data Mining Library.'' \emph{Journal of Open Source Software} 4 (36): 1230.

\leavevmode\hypertarget{ref-propp1996exact}{}%
Propp, James Gary, and David Bruce Wilson. 1996. ``Exact Sampling with Coupled Markov Chains and Applications to Statistical Mechanics.'' \emph{Random Structures and Algorithms} 9 (1-2): 223--52.

\leavevmode\hypertarget{ref-roberts2001optimal}{}%
Roberts, Gareth O, and Jeffrey S Rosenthal. 2001. ``Optimal Scaling for Various Metropolis-Hastings Algorithms.'' \emph{Statistical Science} 16 (4): 351--67.

\leavevmode\hypertarget{ref-robins2007introduction}{}%
Robins, Garry, Pip Pattison, Yuval Kalish, and Dean Lusher. 2007. ``An Introduction to Exponential Random Graph (P*) Models for Social Networks.'' \emph{Social Networks} 29 (2): 173--91.

\leavevmode\hypertarget{ref-rue2001fast}{}%
Rue, Håvard. 2001. ``Fast Sampling of Gaussian Markov Random Fields.'' \emph{Journal of the Royal Statistical Society: Series B (Statistical Methodology)} 63 (2): 325--38.

\leavevmode\hypertarget{ref-rue2005gaussian}{}%
Rue, Håvard, and Leonhard Held. 2005. \emph{Gaussian Markov Random Fields: Theory and Applications}. CRC Press.

\leavevmode\hypertarget{ref-schweinberger2015local}{}%
Schweinberger, Michael, and Mark S Handcock. 2015. ``Local Dependence in Random Graph Models: Characterization, Properties and Statistical Inference.'' \emph{Journal of the Royal Statistical Society: Series B (Statistical Methodology)} 77 (3): 647--76.

\leavevmode\hypertarget{ref-snijders2006new}{}%
Snijders, Tom AB, Philippa E Pattison, Garry L Robins, and Mark S Handcock. 2006. ``New Specifications for Exponential Random Graph Models.'' \emph{Sociological Methodology} 36 (1): 99--153.

\leavevmode\hypertarget{ref-laplacesdemon}{}%
Statisticat, LLC. 2016. \emph{LaplacesDemon: Complete Environment for Bayesian Inference}. Bayesian-Inference.com. \url{https://web.archive.org/web/20150206004624/http://www.bayesian-inference.com/software}.

\leavevmode\hypertarget{ref-strauss1990pseudolikelihood}{}%
Strauss, David, and Michael Ikeda. 1990. ``Pseudolikelihood Estimation for Social Networks.'' \emph{Journal of the American Statistical Association} 85 (409): 204--12.

\leavevmode\hypertarget{ref-swendsen1987nonuniversal}{}%
Swendsen, Robert H, and Jian-Sheng Wang. 1987. ``Nonuniversal Critical Dynamics in Monte Carlo Simulations.'' \emph{Physical Review Letters} 58 (2): 86.

\leavevmode\hypertarget{ref-turek2017automated}{}%
Turek, Daniel, Perry de Valpine, Christopher J Paciorek, and Clifford Anderson-Bergman. 2017. ``Automated Parameter Blocking for Efficient Markov Chain Monte Carlo Sampling.'' \emph{Bayesian Analysis} 12 (2): 465--90.

\leavevmode\hypertarget{ref-wasserman1996logit}{}%
Wasserman, Stanley, and Philippa Pattison. 1996. ``Logit Models and Logistic Regressions for Social Networks: I. An Introduction to Markov Graphs Andp.'' \emph{Psychometrika} 61 (3): 401--25.

\leavevmode\hypertarget{ref-zhang2001segmentation}{}%
Zhang, Yongyue, Michael Brady, and Stephen Smith. 2001. ``Segmentation of Brain Mr Images Through a Hidden Markov Random Field Model and the Expectation-Maximization Algorithm.'' \emph{IEEE Transactions on Medical Imaging} 20 (1): 45--57.

\end{document}